\documentclass{elsart3p}

\usepackage{epsfig}

\usepackage{amssymb}

\begin{document}

\begin{frontmatter}



\title{Altitude dependence of fluorescence light emission by extensive air showers}

\author[Uni]{B.~Keilhauer\corauthref{cor1}}
\author[Uni,FZK]{J.~Bl\"umer}
\author[FZK]{R.~Engel}
\author[FZK]{H.O.~Klages}
\address[Uni]{Universit\"at Karlsruhe, Institut f\"ur Experimentelle Kernphysik,
76021 Karlsruhe, Germany}
\address[FZK]{Forschungszentrum Karlsruhe, Institut f\"ur Kernphysik, 76021
Karlsruhe, Germany}
\corauth[cor1]{bianca.keilhauer@ik.fzk.de}

\begin{abstract}
  Fluorescence light is induced by extensive air showers while
  developing in the Earth's atmosphere. The number of emitted
  fluorescence photons depends on the conditions of the air and on the
  energy deposited by the shower particles at every stage of the
  development. In a previous model calculation, the pressure and
  temperature dependences of the fluorescence yield have been
  studied on the basis of kinetic gas theory, assuming
  temperature-independent molecular collision cross-sections. In this
  work we investigate the importance of temperature-dependent
  collision cross-sections and of water vapour quenching on the
  expected fluorescence yield. The calculations will be applied to
  simulated air showers while using actual atmospheric profiles to estimate 
  the influence on the reconstructed energy of extensive air showers.
\end{abstract}

\journal{5th Fluorescence Workshop, Madrid, 2007}
\end{frontmatter}

% main text
\section{Introduction}
\label{intro}

Several air shower experiments like HiRes~\cite{hires}, the Pierre Auger
Observatory~\cite{auger}, and Telescope Array~\cite{ta}, are using the fluorescence
technique for detecting extensive air showers (EAS) induced by ultra-high energy cosmic rays. 
Measuring the fluorescence light that nitrogen
molecules emit after being excited by charged particles of EAS is currently the most direct method for
determining the energy of EAS in a model-independent way. A thorough understanding of the light emission
process is necessary to obtain the primary energy of EAS with high precision. 

In this paper, we extend our previous model calculation for the
fluorescence light emission~\cite{BK2006} by including the latest
results on input parameters and their temperature dependence as
obtained in laboratory measurements. For the reconstruction of air shower 
events, the light emission has to be known in dependence on
altitude in the Earth's atmosphere at which the shower is observed. Up
to now, the altitude dependence has been considered by including air
density profiles and collisional quenching of nitrogen-nitrogen and
nitrogen-oxygen molecules as described by kinetic gas theory. The
cross-sections for collisional quenching were taken to be temperature
independent. However, the
cross-sections describing collisional quenching are
known to be temperature-dependent~\cite{gs}. Gr\"un and Schopper~\cite{gs}
found a decreasing collisional quenching cross-section with increasing
temperature. Recently, the AirFly experiment has studied
collisional quenching cross-sections in dependence on
temperature~\cite{airfly_icrc}. These data are also included in the
model calculations presented in this article. In addition we
investigate the influence of water vapour on the fluorescence yield,
using relative humidity measurements performed at the site of the
Auger detector in Argentina.

\section{Model calculation and experimental data\label{comp}}

Following the mathematical description in~\cite{BK2006}, the
fluorescence yield (number of photons of wavelength $\lambda$ produced
per meter track length) is written as
\begin{eqnarray}
FY_\lambda ~=~
\frac{\varepsilon_\lambda^0}{1+(p/p_{\nu^\prime}^\prime(T))}\cdot
\frac{\lambda}{hc}\cdot\frac{dE}{dX}\cdot\rho_{air},
\label{eq.FY}
\end{eqnarray}
with $\varepsilon_\lambda^0$ being the fluorescence efficiency at wavelength
$\lambda$ without collisional quenching, $p$ is air pressure, $p^\prime_{\nu^\prime}$ is a
reference pressure at which the mean life time of the radiative transition to any lower
state $\tau_0$ is equal to that of collisional quenching $\tau_c$. The index $\nu^\prime$
indicates the excitation level of a band system. The air density is given
by $\rho_{air}$ and the energy deposited locally by a charged particle of 
an EAS is $\frac{dE}{dX}$. 

Assuming air to be a two-component gas, the relation between $p$ and
$p^\prime_{\nu^\prime}$ is given by
\begin{eqnarray}
\frac{p}{p^\prime_{\nu^\prime}}&=&  \frac{\tau_{0,\nu^\prime}
p_{{\rm air}}\cdot N_A}{R\cdot T}\cdot\sqrt{\frac{kTN_A}{\pi}} \label{eq:pprime}
\end{eqnarray}
\vspace{-24pt}
\begin{eqnarray*}
&~&~\cdot\biggl(4\cdot C_v({\rm N}_2)\cdot
\sigma_{{\rm NN},\nu^\prime}(T)\cdot 
\sqrt{\frac{1}{M_{m,{\rm N}}}}
\end{eqnarray*}
\vspace{-20pt}
\begin{eqnarray*} 
&~&~+~2\cdot
C_v({\rm O}_2)\cdot
\sigma_{{\rm NO},\nu^\prime}(T)
\cdot\sqrt{2(\frac{1}{M_{m,{\rm N}}}+\frac{1}{M_{m,{\rm O}}})}\biggr),  
\end{eqnarray*}
with the masses per mole for nitrogen $M_{m,{\rm N}}$ and oxygen $M_{m,{\rm O}}$ and the
fractional part per volume $C_v$ of the two gas components. 
The temperature dependence of the collisional quenching cross-sections is parametrised as
\begin{equation}
\sigma_{{\rm Nx},\nu^\prime}(T) = \sigma_{{\rm Nx},\nu^\prime}^0\cdot T^{\alpha_{\nu^\prime}}.
\end{equation}
For example, in case of the AirFly experiment, it is given  by $\sigma_{{\rm
Nx},\nu^\prime}^0 = \sigma_{{\rm Nx},\nu^\prime} \cdot 293^{-\alpha_{\nu^\prime}}$. The
cross-sections have been measured for the bands at 313.6~nm (2P(2-1)), 337.1~nm (2P(0-0)),
353.7~nm (2P(1-2)), and 391.4~nm (1N(0-0)). In the calculations presented here, it is
assumed that $\alpha_{\nu^\prime}$ is the same for all bands within its band system. In
the AirFly  experiment, $\alpha_{\nu^\prime}$ have been measured for dry air and no
differentiation has been made for nitrogen or oxygen. For fitting those data into the model
calculation, it is assumed that these $\alpha$-coefficients can be applied to the 
quenching cross-sections of both N-N and N-O collisions. 

An absolute calibration of the AirFly experiment has not yet been
published. Therefore, the fluorescence efficiency
$\epsilon_{337.1nm}^0$ has been set to 0.082\% of the deposited
energy, the value given by Bunner~\cite{bunner}. The same
normalisation for $\epsilon_{337.1nm}^0$ is used in the calculations
of \cite{BK2006}.

In the following we will show two model calculations based on AirFly
measurements. If the temperature dependence of the collisional
quenching cross-sections is not considered, Eq.~(\ref{eq.FY}) is used
with the parameters $p^\prime_{\nu^\prime}$ given in \cite{airfly}.
Eq.~(\ref{eq:pprime}) is applied for calculating the fluorescence yield
with temperature-dependent cross-sections.

\begin{figure}[htbp]
\epsfig{file=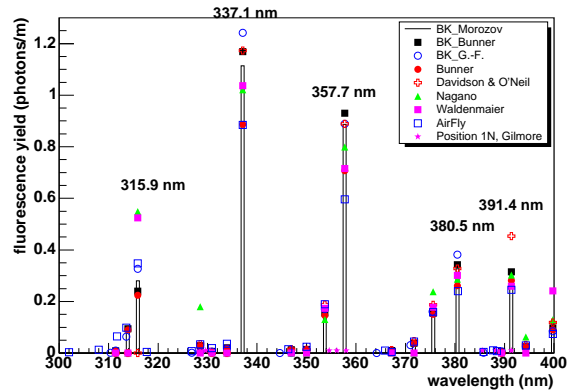,width=0.5\textwidth}
\caption{Fluorescence yield spectra of several calculations and measurements for 0.85~MeV
electrons as exciting particles in the US Standard Atmosphere at sea level. 
The calculations following the mathematical description in~\cite{BK2006} are
labelled with BK\_\emph{name}, where BK stands for the initials of the
corresponding author of~\cite{BK2006} and \emph{name}
indicates the authors of the input parameters used. Details can be found in the text.
The bars indicate the combination of calculation and input parameters which is
favoured in~\cite{BK2006}. \label{fig:autorenvgl}}
\end{figure}

The resulting fluorescence yield
spectrum, calculated for the US Standard Atmosphere at sea level, 
can be seen in Fig.~\ref{fig:autorenvgl} in comparison with calculations and
other measurements presented in detail in~\cite{BK2006}.
The calculations following the mathematical description in~\cite{BK2006} are
labelled with BK\_\emph{name}, where BK stands for the initials of the
corresponding author of~\cite{BK2006} and \emph{name}
indicates the authors of the input parameters used. Input parameters in
Eq.~\ref{eq:pprime} are the deactivation constants which are 
the radiative life time $\tau_{0,\nu^\prime}$ and the collisional cross-section 
between nitrogen and nitrogen molecules $\sigma_{{\rm NN},\nu^\prime}$ and between
nitrogen and oxygen molecules $\sigma_{{\rm NO},\nu^\prime}$. Bunner provides collisional
cross-sections and radiative life times for the most prominent band systems of
nitrogen~\cite{bunner}. Using these input parameters in the calculation, the results are
labelled with BK\_Bunner. Recent measurements by Morozov et al.~\cite{morozov} were 
performed for the 2P $\nu^\prime$ = 0,1 band systems. For results named BK\_Morozov,
the values from Bunner are replaced by the newer data by Morozov et al.~where available. 
An alternative calculation of the fluorescence efficiency without collisional quenching
$\epsilon^0_\lambda$ is also presented in~\cite{BK2006}. Here the Einstein coefficients
$A_{\nu^\prime\nu^{\prime\prime}}$ and the radiative life times from Gilmore et
al.~\cite{gilmore} and the \emph{relative apparent excitation cross-section} $Q_{\rm app}$
from Fons et al.~\cite{fons} are used. For details of this procedure
see~\cite{BK2006}. The resulting fluorescence yield is labelled with BK\_G.-F.

\begin{figure}[htbp]
\epsfig{file=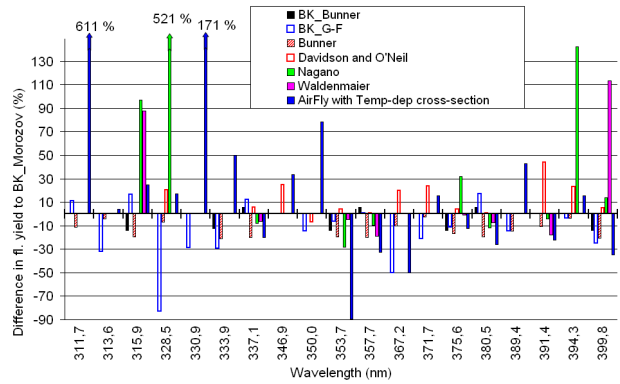,width=0.5\textwidth}
\caption{Relative comparison of 19 bands of the ``BK\_Morozov''-calculation with
measurements and further calculations. The absolute fluorescence 
yield of these contributions can be seen in Fig.~\ref{fig:autorenvgl} with the same colour
labels. \label{fig:relVgl_autoren}}  
\end{figure}

For a relative comparison of 19 bands between 300 and 400~nm see Fig.~\ref{fig:relVgl_autoren}.
The basis is the calculation ``BK\_Morozov''.

\section{Temperature-dependent cross-section\label{temp-dep}}

In the following, the fluorescence yield for 0.85~MeV electrons and for EAS are studied
using the ``BK\_Morozov''-calculation upgraded by temperature-dependent collisional quenching
cross-sections. 

\begin{figure}[htbp]
\epsfig{file=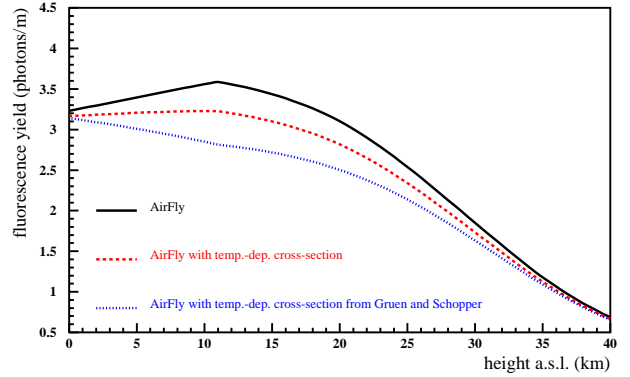,width=0.5\textwidth} 
\caption{Fluorescence yield profiles for a 0.85~MeV electron in
the US Standard Atmosphere with measured parameters from the AirFly
experiment~\cite{airfly,airfly_icrc}. The yield is the sum of 26 bands between 300 and
400~nm as listed in~\cite{airfly}. See
text for details. \label{fig:yield-elec_Paolo_temptest_gs}}  
\end{figure}

The fluorescence yield profiles for a 0.85~MeV electron, corresponding
to an energy deposit of 0.1677~GeV per g/cm$^2$, is shown in
Fig.~\ref{fig:yield-elec_Paolo_temptest_gs} for the US Standard
Atmosphere.
The yield is the sum of 26 bands between 300 and 400~nm as listed
in~\cite{airfly}.  The solid black line represents the results where
no temperature-dependent collisional quenching cross-sections have
been considered. The red dashed line includes the
temperature-dependent cross-section as measured by
AirFly. The blue dotted line reflects the AirFly data where the
$\alpha$-coefficients for the temperature-dependent cross-sections are replaced
by the data from Gr\"un and Schopper~\cite{gs}. The value has been
extracted from Fig.~6 of~\cite{gs} and is therefore quite
imprecise. Additionally, a caveat has to be applied to this comparison: Gr\"un and
Schopper measured one $\alpha$-coefficient for the entire wavelength range 
in pure nitrogen and not in air. For 
this simple comparison, the value for N-N collisions has been used 
for the $p^\prime$ of air. However, the AirFly result confirms a decreasing
cross-section with increasing temperature and also the absolute scales
of the two publications are in quite good agreement.
\begin{figure}[htbp]
\epsfig{file=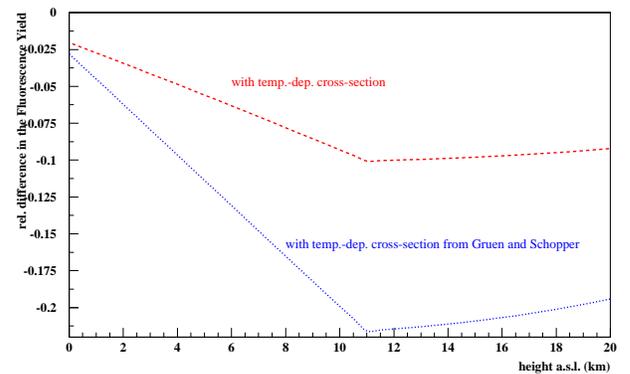,width=0.5\textwidth} 
\caption{Relative difference of the fluorescence yield calculated with
temperature-dependent collisional cross-sections compared with the former model
calculation. \label{fig:delta-yield_Paolo_gs}}  
\end{figure}
The new dependence reduces the fluorescence yield at sea level by about
2\% compared with the former model calculation, see 
Fig~\ref{fig:delta-yield_Paolo_gs}. 
This deficit increases
with increasing altitude to about 10\% at 11~km a.s.l.~in case of
AirFly and to about 20\% in case of Gr\"un and Schopper data.  

\begin{figure}[htbp]
\epsfig{file=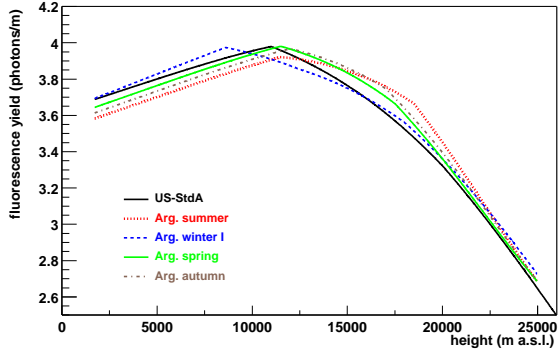,width=0.5\textwidth} 
\caption{Fluorescence yield profiles for a 0.85~MeV electron in the US Standard Atmosphere
and measured Argentine atmospheres as given in~\cite{BK2004}. The given yield is a sum of
all emitted photons between 300 and 400~nm. \label{fig:yield_elec_Arg}}  
\end{figure}

\begin{figure}[htbp]
\epsfig{file=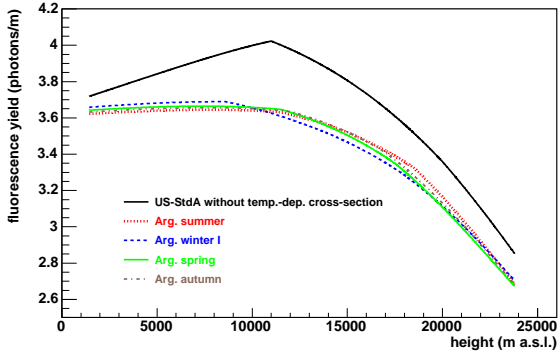,width=0.5\textwidth} 
\caption{Fluorescence yield profiles for a 0.85~MeV electron in
the US Standard Atmosphere and measured Argentine atmospheres as given
in~\cite{BK2004}. The curves are calculated with the ``BK\_Morozov''-model of~\cite{BK2006}
combined with the temperature-dependent collisional quenching cross-sections from
AirFly~\cite{airfly_icrc}.\label{fig:yield_elec_Tempdep_0vapordepMal_USodep}}
\end{figure}

\begin{figure}[htbp]
\epsfig{file=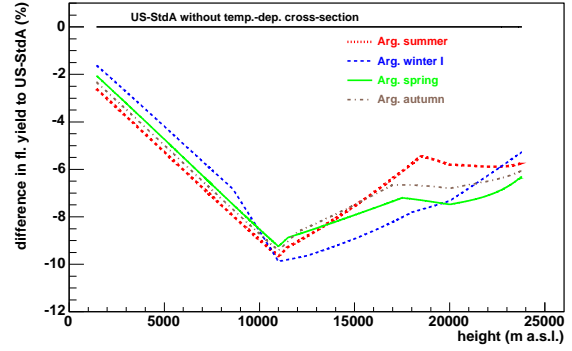,width=0.5\textwidth} 
\caption{Difference of the fluorescence yield profiles for a 0.85~MeV electron in measured Argentine atmospheres to those in the US Standard Atmosphere
without temperature-dependent collisional quenching
cross-sections.\label{fig:dpyield_elec_Tempdep_0vapordepMal_USodep}}  
\end{figure}

In the next step, the calculation of the fluorescence yield is performed using Argentine
atmospheres as given in~\cite{BK2004}. The former model calculation leads to fluorescence
yield profiles for the seasonal atmospheres as can be seen in
Fig.~\ref{fig:yield_elec_Arg}~\cite{BK2006}.
Applying the temperature-dependent collisional cross-sections of AirFly in the calculations using
Argentine atmospheres, the fluorescence yield profiles are strongly distorted, see
Fig.~\ref{fig:yield_elec_Tempdep_0vapordepMal_USodep}. 
The different temperature profiles lead to a varying strength of the influences of the
temperature-dependent cross-sections. In general, it can be stated that the
temperature dependence reduces the increase of the fluorescence yield within the lowest 11~km
in the 
atmosphere significantly. Furthermore, the differences between the seasonal atmospheres
are reduced compared with the calculations without the temperature-dependent cross-sections.
Fig.~\ref{fig:dpyield_elec_Tempdep_0vapordepMal_USodep} displays the difference of the
fluorescence yield in Argentine atmospheres to that in the US Standard Atmosphere with the
former calculation.

\begin{figure}[htbp]
\epsfig{file=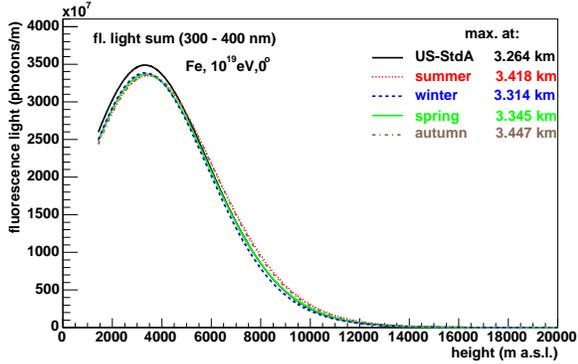,width=0.5\textwidth} 
\caption{Fluorescence light profiles for an iron-induced extensive air shower with $E_0$ =
10$^{19}$~eV and vertical incidence in
the US Standard Atmosphere and measured Argentine atmospheres as given
in~\cite{BK2004}. The fluorescence emission is calculated with the ``BK\_Morozov''-model
of~\cite{BK2006} combined with the temperature-dependent collisional quenching cross-section from
AirFly~\cite{airfly_icrc} and the extensive air shower is simulated with
CORSIKA~\cite{heck}.\label{fig:fl_yield_E19_0deg_wTempdep_USodep}}  
\end{figure}

\begin{figure}[htbp]
\epsfig{file=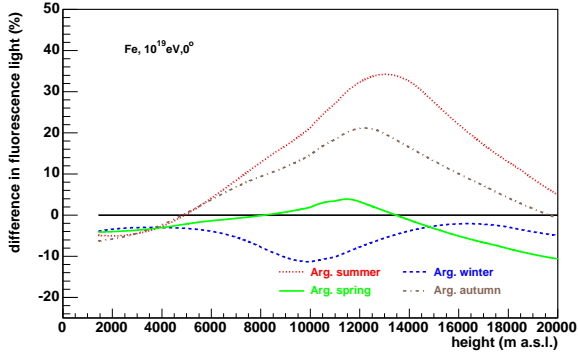,width=0.5\textwidth} 
\caption{Difference of the fluorescence light profiles as shown in
Fig.~\ref{fig:fl_yield_E19_0deg_wTempdep_USodep}.
\label{fig:dpyield_E19_0deg_wTempdep_USodep}} 
\end{figure}

\begin{figure}[htbp]
\epsfig{file=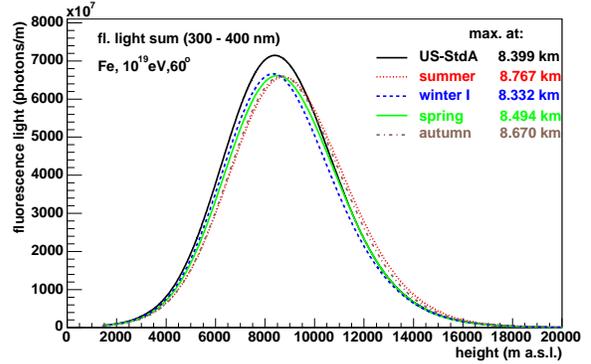,width=0.5\textwidth} 
\caption{Fluorescence light profiles for an iron-induced extensive air shower with $E_0$ =
10$^{19}$~eV and 60$^\circ$ inclination angle. Apart from the inclination, everything is
the same as in
Fig.~\ref{fig:fl_yield_E19_0deg_wTempdep_USodep}.
\label{fig:fl_yield_E19_60deg_wTempdep_USodep}}   
\end{figure}

\begin{figure}[htbp]
\epsfig{file=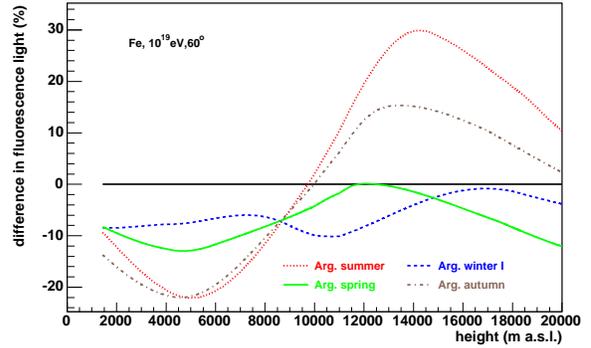,width=0.5\textwidth} 
\caption{Difference of the fluorescence light profiles as shown in
Fig.~\ref{fig:fl_yield_E19_60deg_wTempdep_USodep}.
\label{fig:dpyield_E19_60deg_wTempdep_USodep}} 
\end{figure}

For estimating the importance of temperature-dependent cross-sections
on reconstructing EAS profiles, two average iron-induced EAS with $E_0
= 10^{19}$~eV have been simulated with CORSIKA~\cite{heck}, one with
vertical incidence and the other with 60$^\circ$ inclination. The
simulations have been performed with the US Standard Atmosphere and
afterwards, the conversion from atmospheric depth $X$ to geometric
altitude $h$ has been done applying Argentine
atmospheres~\cite{BK2004}. For these EAS profiles, the fluorescence
light is calculated and shown in
Fig.~\ref{fig:fl_yield_E19_0deg_wTempdep_USodep} for the vertical case
and in Fig.~\ref{fig:fl_yield_E19_60deg_wTempdep_USodep} for the EAS
with 60$^\circ$ inclination.
The corresponding differences of the fluorescence light in Argentine atmospheres to that in
the US Standard Atmosphere are displayed in
Figs.~\ref{fig:dpyield_E19_0deg_wTempdep_USodep} and
\ref{fig:dpyield_E19_60deg_wTempdep_USodep}. 
The primary energy of EAS can be obtained be converting the fluorescence light into local
energy deposit and then integrating over the entire profile of an EAS. Including
temperature-dependent collisional quenching cross-section in the calculations, 
the expected shower light profile of an EAS with vertical incidence is reduced by 2.7\% in
Argentine summer, by 3.2\% in autumn, by 3.7\% in spring, and by 4.8\% in winter. Doing
the same for the 
60$^\circ$ inclined shower, the reduction of the expected light increases to 6.3\%
in summer, 6.8\% in autumn, 7.1\% in spring, and 7.5\% in winter. 
A former study has shown that varying atmospheric profiles influence the longitudinal
shower development. It yields in
slightly distorted profiles of the energy deposit of the air shower, which means an
uncertainty of the energy reconstruction of the primary particle of less than
1\%~\cite{icrc05}. The position of the shower maximum has also been studied and a shift
of about -15 g cm$^{-2}$ on average was found. Applying additionally the
temperature-dependent cross-sections, the position of the shower maximum is only shifted
slightly beyond it. In all atmospheric models, the additional shift is less than
50~m. 

The same calculations have been performed for proton-induced air showers with the same
parameters. The reduction of
the expected light is increased by about 0.5\% compared to the numbers of
the iron-induced showers.

\section{Vapour quenching}
\label{vapor-dep}

\begin{figure}[htbp]
\epsfig{file=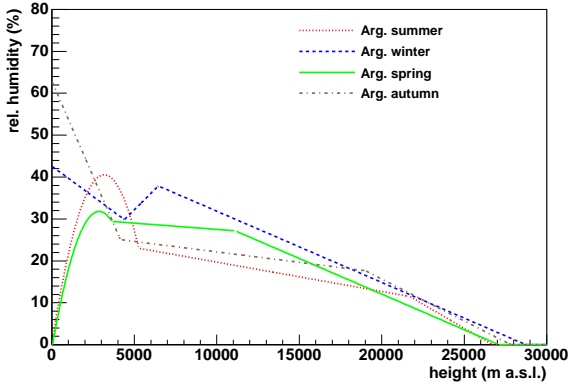,width=0.5\textwidth}
\caption{Seasonal average profiles of relative humidity at the site of the Pierre Auger
Observatory in Argentina as measured during night times. \label{fig:vapor_graphs}}  
\end{figure}

In this study, the model calculation has been expanded by including collisional
quenching due to water vapour. An additional term is inserted in Eq.~(\ref{eq:pprime}) to
account for the collisions between nitrogen and water vapour molecules. The corresponding
cross-section has been measured by e.g.~Morozov et al.~\cite{morozov} and
Waldenmaier~\cite{waldenmaier}. It has to be stressed that no temperature-dependence has
been measured for the collisional quenching cross-section between water vapour and
nitrogen. Typically in all atmospheric models, the humidity is set 
to zero. For this study, seasonal average profiles of relative humidity at the site of the
Pierre Auger Observatory in Argentina, measured during night times, are fitted, see
Fig.~\ref{fig:vapor_graphs}. 

\begin{figure}[htbp]
\epsfig{file=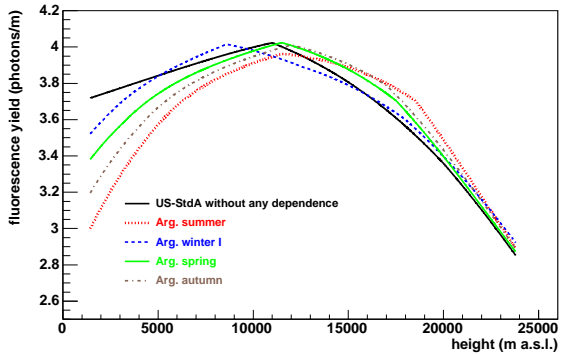,width=0.5\textwidth} 
\caption{Fluorescence yield profiles for a 0.85~MeV electron in
the US Standard Atmosphere and Argentine atmospheres. The fluorescence emission is
calculated with the ``BK\_Morozov''-model of~\cite{BK2006} including vapour quenching using the
humidity profiles given in Fig.~\ref{fig:vapor_graphs}, however no temperature-dependent
collisional quenching cross-section has been
applied. \label{fig:yield_elec_oTempdep_vapordepMal_USodep}}   
\end{figure}

\begin{figure}[htbp]
\epsfig{file=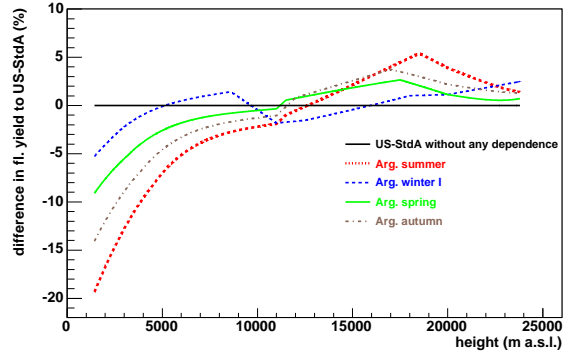,width=0.5\textwidth} 
\caption{Difference of the fluorescence yield profiles in Argentine atmospheres as shown in
Fig.~\ref{fig:yield_elec_oTempdep_vapordepMal_USodep} to the US Standard Atmosphere.
\label{fig:dpyield_elec_oTempdep_vapordepMal_USodep}} 
\end{figure}

Firstly, vapour quenching has been considered in the former model calculations
applying the cross-sections measured by Morozov et al.~\cite{morozov} without
temperature-dependent cross-sections. The resulting fluorescence yield 
profiles for a 0.85~MeV electron can be seen in
Fig.~\ref{fig:yield_elec_oTempdep_vapordepMal_USodep}. 
Vapour quenching reduces the fluorescence yield mainly in the lower part of the
atmosphere, compare with Fig.~\ref{fig:yield_elec_Arg},
because the water vapour content in the atmosphere is low at higher altitudes. The
differences of the calculations in Argentine atmospheres to the US Standard Atmosphere are
shown in Fig.~\ref{fig:dpyield_elec_oTempdep_vapordepMal_USodep}.
The effect is largest during summer, because more water vapour can be
contained in warmer air. The reduction of fluorescence yield is most
significant near the ground, about 20\%, and becomes less than 5\%
above 7~km a.s.l. During winter, the effect is of minor importance and
only visible below about 3~km a.s.l.

\begin{figure}[htbp]
\epsfig{file=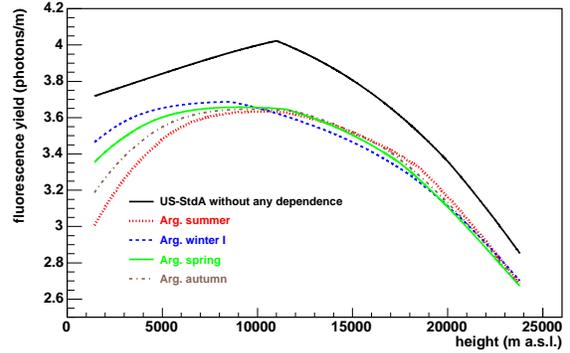,width=0.5\textwidth} 
\caption{Fluorescence yield profiles for a 0.85~MeV electron in
the US Standard Atmosphere and Argentine atmospheres. The fluorescence emission is
calculated with the ``BK\_Morozov''-model of~\cite{BK2006} including vapour quenching using the
humidity profiles given in Fig.~\ref{fig:vapor_graphs} and temperature-dependent cross-sections.
\label{fig:yield_elec_wTempdep_vapordepMal_USodep}}  
\end{figure}

\begin{figure}[htbp]
\epsfig{file=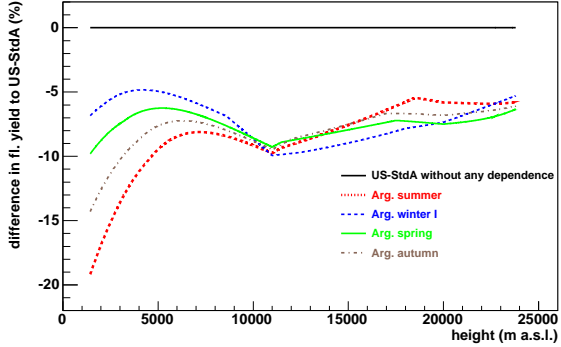,width=0.5\textwidth} 
\caption{Difference of the fluorescence yield profiles as shown in
Fig.~\ref{fig:yield_elec_wTempdep_vapordepMal_USodep}.
\label{fig:dpyield_elec_wTempdep_vapordepMal_USodep}}   
\end{figure}

Secondly, vapour quenching has been included in the model calculation presented in
Sec.~\ref{temp-dep}. Again, the fluorescence yield is determined for a 0.85~MeV electron
in Argentine atmospheres,
Fig.~\ref{fig:yield_elec_wTempdep_vapordepMal_USodep}. 
The difference of the fluorescence yield in Argentine atmospheres to that in the US
Standard Atmosphere can be seen in
Fig.~\ref{fig:dpyield_elec_wTempdep_vapordepMal_USodep}.
The additional vapour quenching changes the profiles shown in
Fig.~\ref{fig:yield_elec_Tempdep_0vapordepMal_USodep} mainly in the lowest part of
the atmosphere. In summer, the fluorescence yield is reduced significantly and in winter
the effect is smallest.

\begin{figure}[htbp]
\epsfig{file=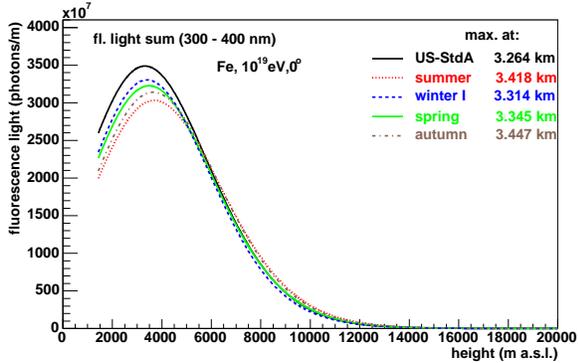,width=0.5\textwidth} 
\caption{Fluorescence light profiles for the iron-induced extensive air shower as shown in
Fig.~\ref{fig:fl_yield_E19_0deg_wTempdep_USodep}. For the fluorescence emission
calculations, additionally the vapour quenching has been
included. \label{fig:fl_yield_E19_0deg_wVapdep_USodep}} 
\end{figure}

\begin{figure}[htbp]
\epsfig{file=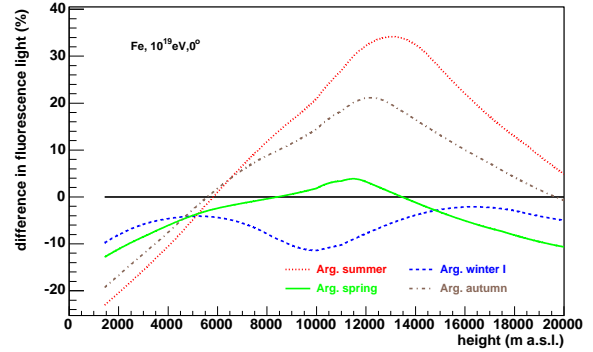,width=0.5\textwidth} 
\caption{Difference of the fluorescence light profiles as shown in
Fig.~\ref{fig:fl_yield_E19_0deg_wVapdep_USodep}.
\label{fig:dpyield_E19_0deg_wVapdep_USodep}}   
\end{figure}

Thirdly, the model calculation including all dependences is applied to the average
iron-induced EAS which are used already in Sec.~\ref{temp-dep}. Since the effect of water
vapour quenching is most important near ground, here only the fluorescence light profiles
are shown for the vertical shower, Fig.~\ref{fig:fl_yield_E19_0deg_wVapdep_USodep}.
The graph
of the differences of the fluorescence light in Argentine atmospheres to that in the US
Standard Atmosphere can be seen in Fig.~\ref{fig:dpyield_E19_0deg_wVapdep_USodep}.
The expected light of EAS is reduced by about additional 8.2\% due to
added water vapour 
quenching in Argentine summer, by 5.5\% during autumn, by 3.4\% during spring, and by
about 2\% during Argentine winter. In total, including the
temperature-dependent collisional cross-sections and the water vapour quenching, the
expected light of the EAS is reduced by 11.1\% during summer, 8.9\% during autumn,
7.3\% during spring, and 6.8\% during winter.

For the 60$^\circ$ inclined shower, the additional effect due to vapour quenching is
smaller and ranges between 1.2\% in summer and 0.2\% in winter. Combining the two
effects, the expected light is reduced by 8.4\% during winter, 8.1\%
during spring and autumn, and 8.0\% during summer.

\section{Conclusion}
\label{concl}

The effects of temperature-dependent collisional quenching
cross-sections and of quenching due to water vapour have been
studied. Both effects lead to a significant reduction of the
fluorescence yield in the lower part of the atmosphere. Applying these
calculations to simulated EAS, a distortion of the longitudinal shower
development is found. 
A reduction of the emitted ligth is expected, which varies from about 7\% to 11\%
depending on seasonal atmospheric model and on zenith angle of the EAS. Hence, accounting
for these effects in the reconstruction of the primary energy of EAS, the primary
energy will be increased by this amount as compared with the former model 
calculations. The
position of the shower maximum is hardly shifted, in all atmospheric
models the shift is less than 50~m.

\subsection*{Acknowledgement}
On of the authors (BK) is supported by the German Research Foundation (DFG) under contract
KE 1151/1-2.

\bibliographystyle{elsart-num}
%\bibliography{biblio}

\end{document}